\documentclass[journal]{IEEEtran}
\usepackage{epsfig}
\usepackage[T1]{fontenc}
\usepackage{lmodern}
\usepackage{textcomp}
\usepackage{latexsym}
\usepackage{cite}
\usepackage{bm}
\usepackage{amsmath}
\usepackage{amsfonts}
\usepackage{multirow} 
\usepackage{url}
\usepackage{subcaption}

\newcommand{\vct}[1]{\mbox{\boldmath{$#1$}}}

\newcommand{\ee}        {\mathrm{e}}
\newcommand{\jj}        {\mathrm{j}}

\begin{document}
\title{Millimeter-Wave Radar Sensing \\of Wombat Respiration}

\author{Marina~Murakami, Ryoko~Iwase, Chiemi~Iba, Daisuke~Ogura, and Takuya~Sakamoto,~\IEEEmembership{Senior Member, IEEE,} 
\thanks{M.~Murakami and T.~Sakamoto are with the Department of Electrical Engineering, Graduate School of Engineering, Kyoto University, Kyoto 615-8510, Japan.}
\thanks{R.~Iwase, C.~Iba, and D.~Ogura are with the Department of Architecture and Architectural Engineering, Graduate School of Engineering, Kyoto University, Kyoto 606-8501, Japan.}}

\maketitle  
  \begin{abstract}
     This study demonstrates the feasibility of radar-based non-contact respiratory monitoring for wombats. Two measurement experiments were conducted in June and December 2024 using 79-GHz millimeter-wave radar systems to monitor the respiration of two wombats. To estimate the respiratory interval, we used a method based on summing harmonic components in the autocorrelation function, capturing the quasi-periodic displacement of the body surface caused by respiration. Estimation accuracy was evaluated through simultaneous measurements from different angles using two radar units. The respiratory interval and respiratory rate were measured with errors of 47.4 ms (2.44\%) and 0.81 bpm (2.21\%), respectively. We also discuss the differences in respiratory rates between the two wombats, as well as seasonal variations between June and December. The results support the potential application of this method to non-contact health monitoring of wombats.
\end{abstract}

  \begin{IEEEkeywords}
    Harmonic component, millimeter-wave radar, non-contact measurement, respiration, wombats.
  \end{IEEEkeywords}
\IEEEpeerreviewmaketitle

\section{Introduction}
There is a growing demand for physiological signal monitoring in medicine and healthcare because such measurements are expected to improve quality of life and overall well-being~\cite{Loncar2019}. This type of monitoring is valuable not only for humans but also for animals~\cite{Abdisa2017}. Although sensor technologies for animal monitoring are rapidly advancing, most conventional methods rely on contact-based sensors, which can cause considerable stress. Attaching such sensors often requires anesthesia or invasive implantable devices, both of which pose health risks. 

A promising alternative is non-contact sensing such as radar-based monitoring. Radar can detect subtle body movements associated with breathing and heartbeats. Because it does not require physical attachment, animals can be monitored continuously over long periods without inducing stress~\cite{Sakamoto2024}. In particular, millimeter-wave radar offers high sensitivity to minute displacements and can selectively detect a target animal even in complex environments. Therefore, radar-based non-contact sensing is a promising approach for physiological sensing of both animals and humans. Radar-based animal monitoring has been already reported for various species such as cats and dogs~\cite{Lai2022,Wang2020, Yoon2023,Tazen2023}, horses~\cite{T_Matsumoto2022}, monkeys~\cite{Minami2023,Sakamoto2023}, and chimpanzees~\cite{Matsumoto2023,Iwata2023}. However, there are still many animal species for which radar monitoring has not been reported yet.

In this study, we focus on common wombats, a marsupial species native to Australia. Some wombat species, such as the northern and southern hairy-nosed wombats, are currently endangered and protected under national conservation efforts. However, their unique behaviors remain poorly understood. Their rounded body shape makes attaching sensors especially challenging. We propose that radar-based non-contact measurement offers a practical and stress-free solution for monitoring their respiration.

We first performed simultaneous measurements on a single wombat using two radar units to evaluate the accuracy of respiratory monitoring. Next, we analyzed individual differences by comparing the respiratory rates of two wombats. Finally, we assessed seasonal variation by measuring the same individual in two different seasons (June and December). To our knowledge, this is the first report of radar-based non-contact physiological monitoring of wombats. This study not only demonstrates the technical feasibility of radar sensing in a new species but also contributes to non-invasive wildlife conservation research.

\section{Radar signal processing}
\IEEEpubidadjcol
In this study, we used a radar system with a multiple-input and multiple-output (MIMO) linear array with three transmitting and four receiving elements, which is approximated by an $N=12$-element virtual linear array. The signal for the $i$th virtual element is $s_i(t,r)$ $(i=0, 1, \cdots, N-1)$, where $t$ is a slow time, and $r$ is a range given as $r=ct'/2$, where $t'$ is a fast time. Next, we generate a complex
radar image $I_\mathrm{c}(t, r, \theta)=\sum_{i=0}^{N-1}\alpha_i w_i(\theta)s_i(t, r)$, where $\alpha_i$ is the weight of a Taylor window, and $w_i(\theta)=\ee^{\jj k d_0 \sin\theta}$ is the complex weight for an azimuthal angle $\theta$ that includes wavenumber $k=2\pi/\lambda$ coresponding to wavelength $\lambda$ and an element spacing $d_0=\lambda/2$ of the virtual array. For the estimated target position $(r, \theta)=(r_0, \theta_0)$, the displacement waveform is obtained as $d(t)=(\lambda/4\pi)\angle I_\mathrm{c}(t, r_0, \theta_0)$.

We define a vector $\vct{d}'(t)=[d(t), d(t+\Delta t), \cdots, d(t+\Delta t(M-1))]^\mathrm{T}$, and then subtract the average using $\vct{d}(t)=\vct{d}'(t)-(1/M)\vct{1}^\mathrm{T}\vct{d}'(t)$ so that the average of the vector elements becomes 0, where $\vct{1}=[1, 1, \cdots, 1]^\mathrm{T}$ is an $M$-dimensional vector. We then calculate the autocorrelation function $\rho_0(t, \tau)=\vct{d}^\mathrm{T}(t)\vct{d}(t+\tau)/|\vct{d}(t)||\vct{d}(t+\tau)|$. If $d(t)$ is periodic, $\rho_0(t, \tau)$ is also periodic in terms of $\tau$. We adopt harmonic summation of the autocorrelation function as 
\begin{equation}
  \rho(t, \tau) = \sum_{\ell=1}^{L}\rho_0(t, \ell\tau)\ee^{-\ell/\ell_0}
\end{equation}
to improve the estimation accuracy of the respiratory rate. Finally, the respiratory interval at time $t$ is estimated as $\hat{\tau}(t)=\arg\max_{\tau}\rho(t, \tau)$, which leads to the estimate of the respiratory rate.

\section{Measurement condition}
In this study, we measured wombat respiratory rates with millimeter-wave radar. The subjects were two elderly males: Wombat A (Fuku/Hector), 20 years old, and Wombat B (Wain), 35 years old, the latter believed to be the oldest wombat in captivity, given the species' average lifespan of ~20 years.
Previous reports state that the normal respiratory rate of wombats ranges from 12 to 16 breaths per minute (bpm) during deep sleep, and from 26 to 32 bpm while dozing under anesthesia during medical examinations~\cite{Elliot2025, Treby2005_2, Hung2020}. However, elevated respiratory rates have also been reported in pathological conditions, such as 40 bpm in cases where disease was suspected~\cite{NSW2020} and up to 72 bpm in a wombat undergoing medical treatment~\cite{Hung2020}.

The radar module used in this study, T14RE\_01080108\_2D (S-Takaya Electronics Industry Co., Ltd., Okayama, Japan), is a millimeter-wave frequency-modulated continuous wave (FMCW) radar with a center frequency of 79 GHz and a bandwidth of 3.6 GHz. It has an equivalent isotropic radiated power (EIRP) of 24 dBm, and features a MIMO array with a 12-channel configuration. The transmitting beamwidth is 8.0$^\circ$ $\times$ 66.0$^\circ$, while the receiving beamwidth is 8.0$^\circ$ $\times$ 90.0$^\circ$, and the range resolution is 44.7 mm.

Measurements were conducted once each in summer (June) and winter (December) at Satsukiyama Zoo in Ikeda, Osaka, Japan. The maximum temperature was approximately 28$^\circ$C and 10$^\circ$C in the summer and winter measurements, respectively.  

\section{Results of respiratory measurement}
\subsection{Accuracy Evaluation}
\begin{figure}[tb]
  \centering
  \begin{minipage}[b]{0.7\columnwidth}
    \centering
     \includegraphics[width=1.0\linewidth]
 {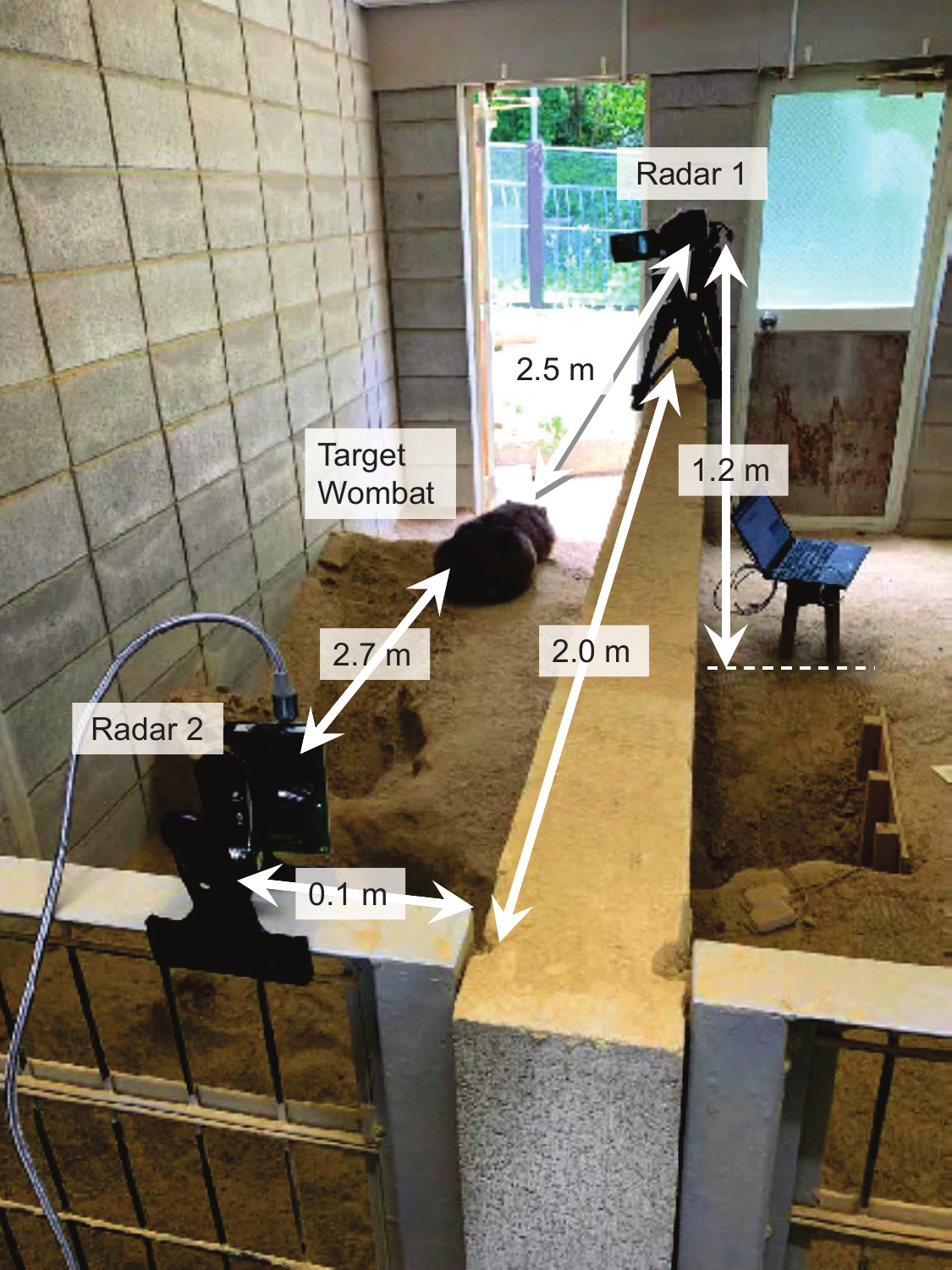}
    \subcaption{}
      \label{fig1a}
  \end{minipage}
  \\\vspace{1mm}\begin{minipage}[b]{1.0\columnwidth}
    \centering
    \includegraphics[width=1.0\linewidth]{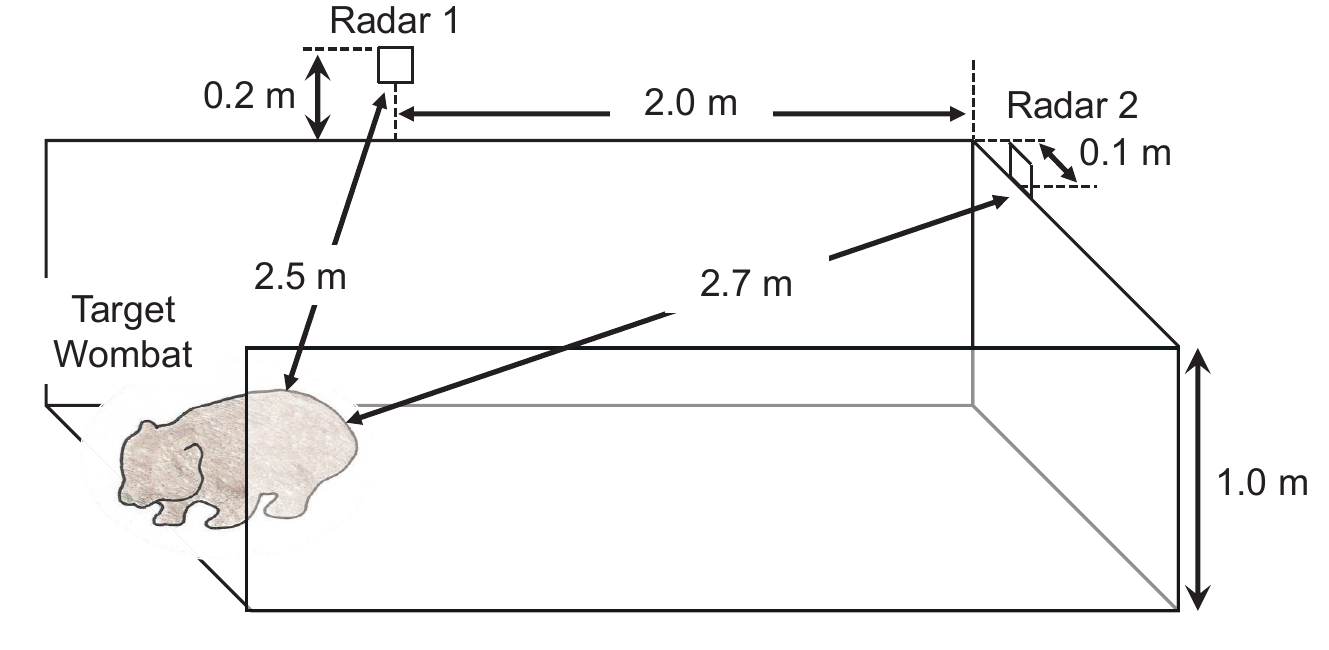}
    \vspace{-8mm}
    \subcaption{}
     \label{fig1b}
\end{minipage}
\caption{(a) Scene of the experiment and (b) schematic of simultaneous measurement by two radars for wombat A}
 \label{fig1}
\end{figure}

Wombat A was simultaneously measured from two angles using two identical radar modules (referred to as radars 1 and 2), as illustrated in Fig.~\ref{fig1} (a) and (b). Wombat A lay low on the ground during the measurement. Radar 1 captured respiratory displacement from the right side of the abdomen, while radar 2 captured it from the lower back area. The measurement was conducted over an 18-minute period,  from 11:37 AM to 11:54 AM.

\begin{figure}[tb]
\centering
    \begin{minipage}[b]{1.0\columnwidth}
    \centering
    \includegraphics[width=0.7\linewidth]{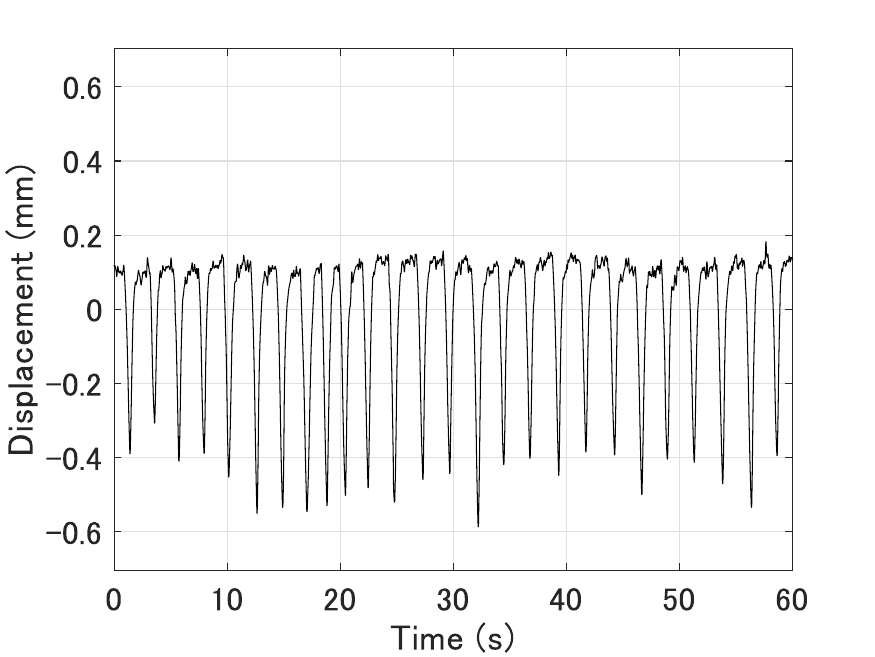}
    \subcaption{Radar 1}
    \end{minipage}
    \begin{minipage}[b]{1.0\columnwidth}
    \centering
    \includegraphics[width=0.7\linewidth]{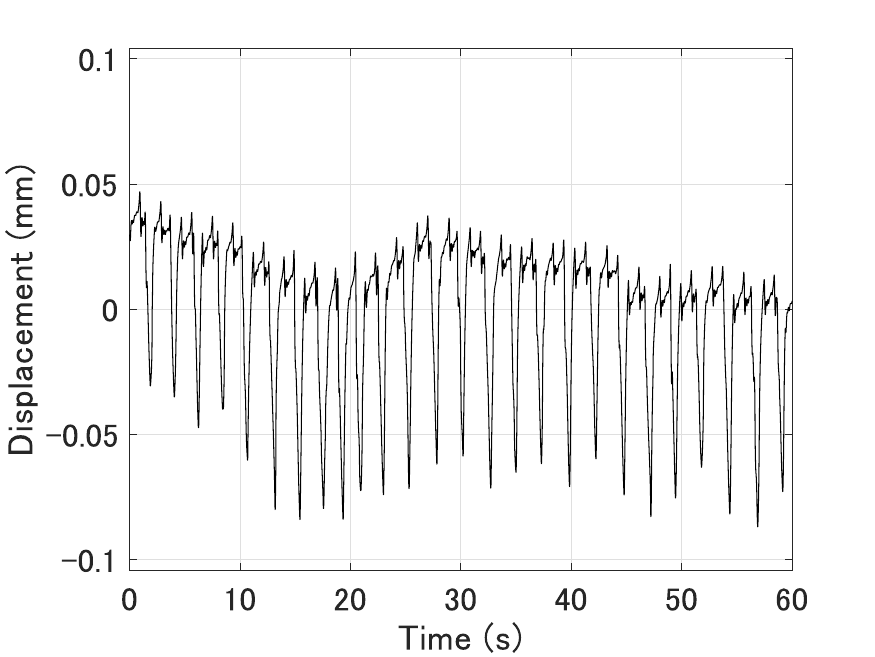}
    \subcaption{Radar 2}
    \end{minipage}
    \caption{Displacements measured using two radars for wombat A.}
      \label{fig3}
\end{figure}

\begin{figure}[tb]
\centering
      \includegraphics[width=0.7\linewidth]{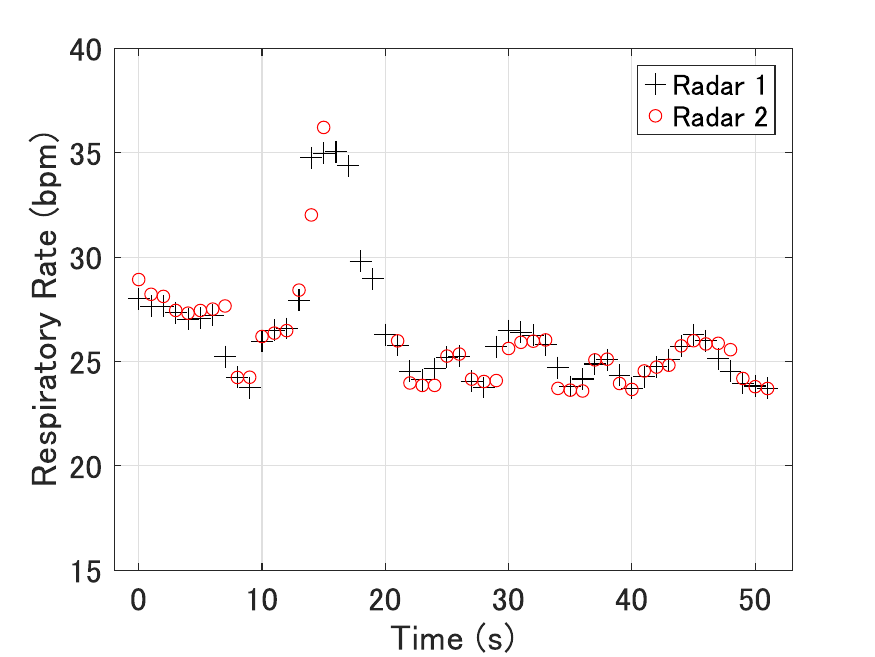}
      \caption{Respiratory rate estimated from two radars for wombat A.}
        \label{fig4}
\centering
\end{figure}

As examples, Fig.~\ref{fig3} shows the measured body displacements, and Fig.~\ref{fig4} presents the estimated respiratory rate during a one-minute interval from 11:43 AM to 11:44 AM. As shown in Fig.~\ref{fig3}, both radar systems captured almost periodic displacements corresponding to respiration. However, the waveforms differ because each radar detected motion from a different part of the body. In particular, radar 2 measured displacements near the wombat's tail, with an amplitude of only $0.1$~mm, which is much smaller than the $0.5$~mm displacement observed in the abdominal region by radar 1. This difference can be explained by anatomical factors: the tail is not only farther from the lungs than the abdomen, but also covered with especially thick skin, which serves to protect the animal from external threats~\cite{encyclopaedia1986}.

\begin{figure}[tb]
\centering
      \includegraphics[width=0.7\linewidth]{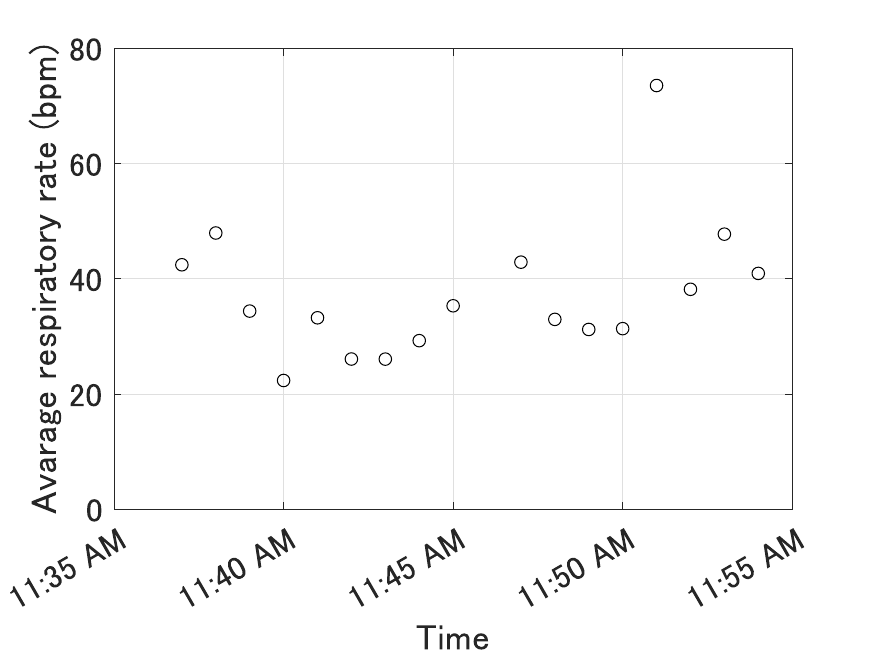}
      \caption{Average respiratory rate over 18 minutes for wombat A.}
        \label{fig5}
\end{figure}

Next, the average respiratory rate for each one-minute interval is shown in Fig.\ref{fig5}. Except for one outlier, the respiratory rate fluctuated between 20 bpm and 50 bpm over the 18-minute measurement period. During the measurement, wombat A remained mostly motionless, lying low on the ground. However, we were unable to estimate the respiratory rate for the one-minute interval starting at 11:46 AM because wombat A moved and changed his position during that time. In addition, brief micro body movements were detected around 11:52 AM and 11:54 AM.

Table \ref{average_Fuku_FA1_WB2} compares the respiratory intervals and respiratory rates measured using radars 1 and 2 over an 18-minute period, from 11:37 AM to 11:54 AM. To reduce the influence of outliers that do not follow a normal distribution, we excluded samples with squared errors in respiratory rate exceeding $3.7$ bpm, which corresponds to $10$\% of the average respiratory rate ($37.4$ bpm) over the entire 18-minute period. We note that the measurement at 11:51 AM corresponds to a relatively high respiratory rate, as also shown in Fig.~\ref{fig5}. 

The mean error of the respiratory interval between the two radars was $47.4$ ms with a root mean square (RMS) error of $2.44$\%. The mean error of the respiratory rate was $0.81$ bpm with an RMS error of $2.21$\%. This will form an important part of our future research on developing a method to suppress outliers. 
The minimum one-minute average respiratory rate recorded in Table \ref{average_Fuku_FA1_WB2} was $22.4$ bpm, which occurred during a resting phase in the first half of the measurement period. This value is consistent with previously reported data for resting wombats~\cite{Elliot2025, Treby2005_2}.
However, the maximum one-minute average respiratory rate was $73.6$ bpm, which is approximately double the values reported previously~\cite{Elliot2025, Treby2005_2}. Because this measurement was taken under natural daily-life conditions without anesthesia or medical intervention, it may represent a novel observation and new insight into wombat physiology.
Furthermore, for the specific one-minute segment shown in Figs.~\ref{fig3} and \ref{fig4}, the RMS errors between radars 1 and 2 were $58.8$ ms for the respiratory interval and 0.74 bpm for the respiratory rate. These are sufficiently low errors for practical purposes. These results demonstrate the effectiveness and accuracy of the proposed non-contact respiratory monitoring method.

\begin{table}
  \caption{Respiratory rate and intervals estimated using radars 1 and 2 for wombat A.}
  \label{average_Fuku_FA1_WB2}
  \centering
\scriptsize
\begin{tabular}{|c|c|c|c|c|c|c|c|c|c|c|}
  \hline
  \multirow{4}{*}{Time} & \multicolumn{2}{c|}{Average} & \multicolumn{2}{c|}{RMS error} & \multicolumn{2}{c|}{Relative error}\\ \cline{2-7}
                        & Resp. & Resp. & Resp. & Resp. & Resp. & Resp.\\
                        & interval    & rate        & interval    & rate        & interval    & rate\\
                        & (s)     & (bpm)      & (ms)    &(bpm)   & (\%)     &(\%) \\ \hline\hline
                 11:37 & $1.75$  & $42.5$     & $35.7$  & $0.43$ & $2.04$   & $1.01$\\ \cline{1-7}
                 11:38 & $1.73$  & $48.0$     & $12.9$ & $0.14$  & $0.74$   & $0.29$\\ \cline{1-7}
                 11:39 & $2.22$  & $34.4$     & $100.4$  & $1.03$  & $4.53$   & $3.00$\\ \cline{1-7}
                 11:40 & $2.87$  & $22.4$     & $38.1$ & $0.27$  & $1.33$   & $1.19$\\ \cline{1-7}
                 11:41 & $2.02$  & $33.2$     & $49.8$ & $0.63$  & $2.47$   & $1.89$\\ \cline{1-7}
                 11:42 & $2.33$  & $26.1$     & $31.4$ & $0.32$  & $1.35$   & $1.23$\\ \cline{1-7}
                 11:43 & $2.32$  & $26.1$     & $58.8$ & $0.74$  & $2.54$   & $2.83$\\ \cline{1-7}
                 11:44 & $2.21$  & $29.3$     & $77.6$ & $0.86$  & $3.51$   & $2.93$\\ \cline{1-7}
                 11:45 & $2.16$  & $35.3$     & $79.0$ & $0.95$  & $3.66$   & $2.69$\\ \hline\hline
                 11:47 & $1.45$  & $42.9$     & $34.6$ & $1.25$  & $2.39$   & $2.91$\\ \cline{1-7}
                 11:48 & $1.83$  & $33.0$     & $46.8$ & $0.81$  & $2.55$ & $2.44$\\ \cline{1-7}
                 11:49 & $1.96$  & $31.2$     & $58.9$ & $0.90$  & $3.01$ & $2.90$\\ \cline{1-7}
                 11:50 & $1.97$  & $31.4$     & $48.0$ & $0.84$  & $2.44$ & $2.68$\\ \cline{1-7}
                 11:51 & $1.26$  & $73.6$     & $14.2$ & $0.99$  & $1.13$ & $1.34$\\ \cline{1-7}
                 11:52 & $1.73$  & $38.2$     & $32.0$ & $0.73$  & $1.85$ & $1.91$\\ \cline{1-7}
                 11:53 & $1.44$  & $47.8$     & $49.3$ & $1.91$  & $3.43$ & $4.00$\\ \cline{1-7}
                 11:54 & $1.52$  & $41.0$     & $37.5$ & $0.93$  & $2.47$ & $2.26$\\ \hline\hline
                 Mean  & $1.93$  & $37.4$     & $47.4$ & $0.81$  & $2.44$ & $2.21$\\ \hline
\end{tabular}
\end{table}

\subsection{Individual and Seasonal Variation}
Measurements were conducted for both wombats A and B in June and wombat B again in December. Both wombats were observed to rest or sleep and breathe naturally, although some partial body movements were noted. In Fig. 5, each bar represents the mean respiratory rate over a 1-min interval, estimated from the radar signals, with error bars indicating the root mean square error. Each group contains five bars, corresponding to five separate 1-min segments selected from periods during which the radar measurements remained stable. From this figure, we can observe differences in respiratory rate that appear to be associated with individual variation as well as seasonal changes. In the June measurements, wombat A exhibited a respiratory rate of approximately 20--30 bpm, while wombat B's was a higher rate of approximately 40--50 bpm. However, in the December measurement, wombat B's respiratory rate was approximately 20 bpm, much lower than the value recorded in June. Wombats lack evaporative cooling mechanisms such as sweating, but stabilize body temperature through behaviors like burrow use and nocturnal activity \cite{Morris2024,Descovich2016}. Increased respiratory rate in summer may also aid heat dissipation, and our measurements suggest this could contribute to resting thermoregulation. Further studies combining thermal environment and body temperature data are needed.

\begin{figure}[tb]
\centering
    \includegraphics[width=0.7\linewidth]{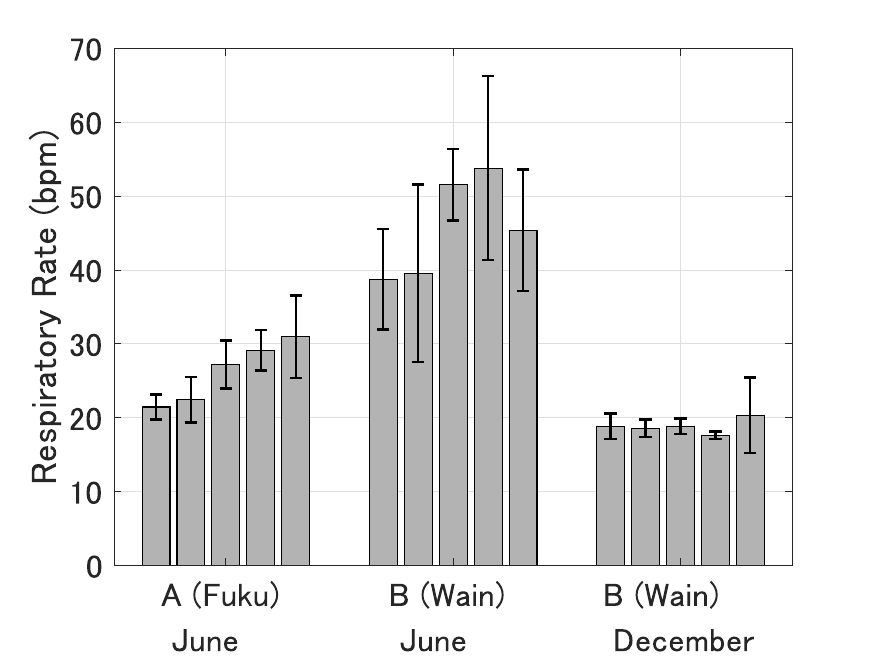}
    \caption{Individual and seasonal variations in respiratory rate.}
      \label{fig6}
\end{figure}

\section{Conclusion}
In this study, we experimentally demonstrated the feasibility of non-contact respiratory measurement for wombats using millimeter-wave radar. First, we conducted simultaneous measurements on a single wombat using two radars and evaluated the measurement accuracy. The results showed that the proposed method can achieve high accuracy with average errors of 47.4 ms in respiratory intervals and 0.81 bpm in respiratory rate estimation. One contributing factor to this high accuracy is the relatively low activity level of wombats, which reduces body movement artifacts during radar measurement. Furthermore, we observed differences in respiratory rates between two individual wombats, as well as seasonal changes by measuring the same individual in both June and December. These findings suggest that radar-based non-contact monitoring can be used to assess not only individual differences but also temporal variations in physiological signals.

Overall, radar-based non-contact measurement represents a promising approach for elucidating the still largely unknown physiology and ecology of wombats. In the future, this technology is expected to contribute to health monitoring, habitat management, and animal welfare not only for wombats but also for other species. At a time when the importance of animal welfare and co-existence between humans and animals is increasingly recognized, this approach has the potential to play a significant role. Therefore, we believe that continuous long-term monitoring using this technology will be of great value moving forward.

\section*{Acknowledgment}
This study involved the installation of sensors within animal enclosures without attaching any devices directly to the animals. As such, it does not constitute animal experimentation. Accordingly, the Kyoto University Animal Experiment Committee determined that formal ethical review was not required.
This work was supported in part by the SECOM Science and Technology Foundation; in part by the Japan Society for Promotion of Science KAKENHI under Grants 21H03427, 23H01420, 23K26115; the New Energy and Industrial Technology; MaRI Co.; the Japan Science and Technology Agency under Grants JPMJMI22J2 and JPMJMS2296. We thank all members of Satsukiyama Zoo, Ikeda City, Osaka, Japan, Mr. Toshiya Suwa of Nikken Sekkei Ltd., and Mr. Kaoru Endo of studio IWASE for their assistance.

\end{document}